\documentstyle[11pt]{article}
\title{Chiral Perturbation Theory and Nucleon Polarizabilities$^*$}
\author{Barry R. Holstein\\
Department of Physics and Astronomy\\
University of Massachusetts\\
Amherst, MA  01003}
\begin{document}
\begin{titlepage}
\maketitle
\begin{abstract}
Compton scattering offers in principle an intriguing new window on
nucleon structure.  Existing experiments and future programs are
discussed and the state of theoretical understanding of such
measurements is explored.
\end{abstract}
\vfill
$^*$Research supported in part by the National Science Foundation
\end{titlepage}
\section{Introduction}
One of the attractive features about low energy Compton scattering from
hadronic systems is that one can make contact with the meaning of such
measurements within the context of classical physics.  This has the
not insignificant consequence that you can explain to your friends
outside particle/nuclear physics what you are doing and why it is of
interest!  The basic idea here is that of {\it polarizability}---{\it
i.e.} the deformation induced in a system in the presence of a
quasistatic electric/magnetic field.\cite{1}  Thus in the presence of an
electric field $\vec{E}_0$ a system of charges will deform and an
electric dipole moment $\vec{p}$ will result.  The electric
polarizability $\alpha_E$ is simply the constant of proportionality between the
applied field and the induced dipole moment
\begin{equation}
\vec{p}=4\pi\alpha_E\vec{E}_0
\end{equation}
Similarly in the presence of a magnetizing field $\vec{H}_0$ a
magnetic dipole moment $\vec{m}$ is generated, with the
proportionality characterized by the magnetic polarizability $\beta_M$
\begin{equation}
\vec{m}=4\pi\beta_M\vec{H}_0
\end{equation}
Obviously then the polarizabilities are fundamental properties of the
hadronic system and probe its underlying structure.

In thinking of how to measure such properties for an elementary
particle it is useful to think initially of a simple atomic system such as
a hydrogen atom.  Then for each such atom there is generated an
energy shift
\begin{equation}
\delta U=-{1\over 2}4\pi\alpha_EE_0^2-{1\over 2}4\pi\beta_MH_0^2\label{eq:aa}
\end{equation}
due to the interaction of the dipole with the fields.\cite{1}
Imagining a box filled with a gas characterized by $N$ atoms per unit
volume, the energy per unit volume of the system of fields plus
atoms will be given by
\begin{eqnarray}
u&=&{1\over 2}(E_0^2+H_0^2)-{N\over
2}(4\pi\alpha_EE_0^2+4\pi\beta_MH_0^2)\nonumber\\
&\equiv&{1\over 2}\epsilon E^2+{1\over 2}\mu H^2
\end{eqnarray}
where
\begin{equation}
E=E_0(1-N4\pi\alpha_E),\qquad H=H_0(1-N4\pi\beta_M)
\end{equation}
are the effective fields in the gas and
\begin{equation}
\epsilon=1+N4\pi\alpha_E,\qquad \mu=1+N4\pi\beta_M
\end{equation}
are the dielectric constant, magnetic permeability respectively.
Using the expression
\begin{equation}
n=\sqrt{\epsilon\mu}=1+N2\pi(\alpha_E+\beta_M)\label{eq:cc}
\end{equation}
which relates the index of refraction $n$ to the dielectric constant
and magnetic permeability we see that measurement of $n$
for our hypothetical gas would provide a sensitive probe for
the sum of electric and magnetic polarizabilities of its 
individual constituents.

In our case, however, we wish to detect the polarizabilities of
an elementary particle---in particular a neutron or proton---and such
an index of refraction experiment is not feasible.  Nevertheless a
means by which to perform such a measurement is suggested by an
alternative way by which to express the index of refraction---in terms
of the forward Compton scattering amplitude $f_k(0)$\cite{2}
\begin{equation}
n=1+N{2\pi\over \omega^2}f_k(0)\label{eq:bb}
\end{equation}
The connection with the polarizability can be made by use of quantum
mechanics.  At lowest order for a charged particle one has the
Hamiltonian
\begin{equation}
H={1\over 2m}(\vec{p}-e\vec{A})^2
\end{equation}
which leads to the well-known Thomson amplitude
\begin{equation}
{\rm Amp}_{\rm Comp}=-{e^2\over m}\hat{\epsilon}\cdot\hat{\epsilon}^{'*}
\end{equation}
for Compton scattering.  Adding on components of the Hamiltonian
corresponding to the polarizabilities---Eq. \ref{eq:aa}---one finds
the modified Compton amplitude
\begin{equation}
{\rm Amp}_{\rm Comp}=\hat{\epsilon}\cdot\hat{\epsilon}^{'*}
(-{e^2\over m}+4\pi\alpha_E\omega\omega')+\vec{k}\times\hat{\epsilon}
\cdot\vec
{k}'\times\hat{\epsilon}^{'*}4\pi\beta_M
\end{equation}
In the forward direction then one has
\begin{equation}
{\rm Amp}_{\rm Comp}(\theta=0)=\hat{\epsilon}\cdot\hat{\epsilon}'
\left[-{e^2\over m}+4\pi(\alpha_E+\beta_M)\omega^2\right]=4\pi f_k(0)
\end{equation}
Then for a {\it neutral} system, we have from Eq. \ref{eq:bb}
\begin{equation}
n=1+N2\pi(\alpha_E+\beta_M)
\end{equation}
in agreement with Eq. \ref{eq:cc}.  However, we now have a
procedure---Compton scattering---which enables the general extraction of the
polarizabilities of an elementary system.  Indeed, calculating the
cross section we find in general
\begin{eqnarray}
{d\sigma\over d\Omega}&=&{\alpha^2\over m^2}\left({\omega '\over
\omega}\right)^2\left\{{1\over 2}(1+\cos^2\theta)
+{m\over \alpha}\omega\omega '[{1\over
2}(\alpha_E+\beta_M)(1+\cos\theta)^2\right.\nonumber\\
&+&\left.{1\over
2}(\alpha_E-\beta_M)(1-\cos\theta)^2]+{\cal O}(\omega^4)\right\}
\end{eqnarray}
so that by measurement of the angular distribution one can extract $\alpha_E,\beta_M$
experimentally.  This program has been carried out for the proton at
SAL and MAMI, yielding (here and below all numerical values for
polarizabilities will be quoted in the units $10^{-4}$ fm$^3$)\cite{3}
\begin{equation}
\alpha_E^p=12.1\pm 0.8\pm 0.5,\qquad \beta_M^p=2.1\mp 0.8\mp 0.5\label{eq:zz}
\end{equation}
In the case of the neutron experiments involving the deuteron are
presently underway at both SAL and MAMI, 
but the best existing number comes from the analysis of a
transmission experiment involving neutron scattering on Pb.  The idea
here is that the existence of a charged particle polarizes the neutron,
which then acts back on the charged particle, generating a $1/r^4$
interaction.  This leads to a term linear in $k$ in a transmission cross
section which can be extracted via careful measurment of its energy
dependence.  The quoted numbers which arise thereby are\cite{4}
\begin{equation}
\alpha_E^n=12.6\pm 1.5\pm 2.0,\qquad \beta_M= 3.2\mp 1.5\mp 2.0\label{eq:yy}
\end{equation}
although the quoted uncertainties are almost certainly too low.\cite{5}

An important contraint in these measurements (and the reason that
errors in the case of the magnetic polarizability are accompanied by
$\mp$) arises from unitarity and causality---{\it i.e.} the feature
that in the forward direction the Compton scattering amplitude can be
represented in terms of a disperson relation involving the total
photoabsorption cross section.  Using a single subtraction, as
indicated from Regge arguments, we have 
\begin{eqnarray}
{\rm Re}f_1(\omega)&=&-{e^2\over m}+{\omega^2\over
2\pi}\int_0^\infty{d\omega'\sigma_{\rm tot}(\omega')\over
{\omega'}^2-\omega^2}\nonumber\\
i.e. && \alpha_E+\beta_M={1\over 2\pi^2}\int_0^\infty
{d\omega'\over {\omega'}^2}\sigma_{\rm tot}(\omega')
\end{eqnarray}

If we include target and beam polarization, things become
more interesting.  Writing, again in the forward direction,
\begin{equation}
{\rm Amp}_{\rm
Comp}=4\pi\left[f_1(\omega)\hat{\epsilon}\cdot\hat{\epsilon}^{'*}+i\omega
f_2(\omega)\vec{\sigma}\cdot\hat{\epsilon}^{'*}\times\hat{\epsilon}\right]
\end{equation}
then the corresponding dispersion relation for $f_2(\omega)$ is
expected to be {\it unsubtracted}!  There exists also in this case a
low energy theorem, first given by Gell-Mann, Goldberger and Low
in terms of the anomalous magnetic moment of the target.\cite{6}  Thus we
write
\begin{equation}
f_2(\omega)=-{e^2\kappa^2\over 2m^2}+\gamma\omega^2+{\cal O}(\omega^4)
\end{equation}
where $\gamma$ is the ``spin polarizability,'' whose relation to the
classical properties of the nucleon is a bit more obscure than in the
case of its unpolarized analogs, but which can be related in a handwaving
fashion to the Faraday effect.  Defining $\sigma_\pm(\omega)$ as the
photoabsorption cross sections with parallel, antiparallel spin and
target helicities, the corresponding dispersion relation yields\cite{7}
\begin{eqnarray}
{\pi e^2\kappa^2\over 2m^2}&=&\int_0^\infty {d\omega\over
\omega}(\sigma_+(\omega)-\sigma_-(\omega))\nonumber\\
\gamma&=&{1\over 4\pi^2}\int_0^\infty{d\omega\over \omega^3}
(\sigma_+(\omega)-\sigma_-(\omega))
\end{eqnarray}
Here the first expression is the Drell-Hearn-Gerasimov sum rule, while
the second provides a dispersive probe of the spin polarizability.

A number of challenges on the experimental front remain
\begin{itemize}
\item [i)] more precise determination of the neutron polarizability,
either by repeating the ORNL measurement or via $d(\gamma,\gamma)$
studies.
\item [ii)] accumulating experimental data utilizing polarization in
order to check the DHG and spin polarizability sum rules.\footnote{It
should be noted in this regard that a possible problem already exists
in that if one looks at the isovector component of the DHG sum rule
one finds
\begin{equation}
{\pi e^2\kappa_n\kappa_p\over m^2}=+15\mu b
\end{equation}
vs.
\begin{equation}
-\int_0^\infty{d\omega\over
\omega}(\sigma_-(\omega)-\sigma_+(\omega))=-39 \mu b
\end{equation}
where the dispersive input has been provided by a multipole analysis
of existing single pion photoproduction data and model-dependent
assumptions about the multipion production.}
\item [iii)] extending the existing measurements in the regime of
virtual Compton scattering---$N(e,e'\gamma)N$---in order to provide a
probe of the local polarizability structure.
\item [iv)] for later use it should be noted that use of the single pion
photoproduction multipole analysis yields a predicted spin
polarizability\cite{8}
\begin{equation}
\gamma\approx -1\times 10^{-4}{\rm fm}^{4}\label{eq:qq}
\end{equation}
\end{itemize}

This then is as far as one can go by means of essentially model independent
analysis.  In the next section we address the question of how well
existing theoretical pictures of the nucleon can confront present and
future measurements.

\section{Theoretical Approaches}

Of course, in addition to having a basic grasp of the
underlying physics it is important to attempt a theoretical
understanding of the nucleon system and its relation to Compton
physics.  In this regard, a first approach which one might employ is
that of a simple constituent quark model.  The idea here is that one
can use well-known sum rules for the electric and magnetic
polarizabilities\cite{9}
\begin{eqnarray}
\alpha_E&=&{\alpha\over 3m}<r_p^2>+2\alpha\sum_{n\neq
0}{|<n|\sum_ie_iz_i|0>|^2\over E_n-E_0}\nonumber\\
\beta_M&=&-{\alpha\over 2m}<(\sum_ie_i\vec{r_i})^2>-{\alpha\over
6}<\sum_ie_i^2{r_i^2\over m_i}>\nonumber\\
&+&2\alpha\sum_{n\neq
0}{|<n|\sum_ie_i{\sigma_{iz}\over 2m_i}|0>|^2\over E_n-E_0}
\end{eqnarray}
Then in a simple harmonic oscillator model of the nucleon there are
only two parameters---the quark mass and the oscillator frequency.
The former is determined in terms of the nucleon mass---$m=M/3$, while the
latter is fixed by the proton charge radius---$\omega=3/M<r_p^2>$.
There exist a number of problems with this approach
\begin{itemize}
\item [i)] The basic scale of the polarizability is too
large---$\alpha_E=2\alpha/M\omega^2\approx 35.$\cite{9}
\item [ii)] The proton electric polarizability is predicted to be
significantly larger than that of the neutron\cite{10}
\begin{equation}
\alpha_E^p-\alpha_E^n\simeq{\alpha\over 3M}<r_p^2>\approx 3.8
\end{equation}
in contradiction to the central values given in Eqs. \ref{eq:zz},\ref{eq:yy}
\item [iii)] There exists a large contribution to the magnetic
polarizability from the $\Delta(1232)$ intermediate
state---$\beta_M^\Delta\approx 12$---which must somehow be cancelled
by an equally large diamagnetic term.\cite{11}
\end{itemize}
At least some of these problems are cured by use of a cloudy bag model
with its intrinsic pion cloud and this suggests that perhaps a better
approach might be the use of chiral perturbative techniques right from
the start, which we next review.

The lowest order chiral Lagrangian coupling pions and nucleons can be
written as
\begin{equation}
{\cal L}=\bar{\Psi}(i\not\!\!{D}-M+{1\over 2}g_A\not\!\!{u}\gamma_5)\Psi\label{eq:pp}
\end{equation}
where (to lowest chiral order) $g_A$ is the nucleon axial decay
constant and $M$ is the nucleon mass.  The coupling to pions is
provided by
\begin{equation}
U=u^2=\exp({i\over F_\pi}\vec{\tau}\cdot\vec{\phi}_\pi)
\end{equation}
while the vector field $u_\mu$ is given by
\begin{equation}
u_\mu=iu^\dagger\nabla_\mu Uu^\dagger
\end{equation}
The covariant derivative $D_\mu\Psi=\partial_\mu\Psi+\Gamma_\mu\Psi$
is given by the connection
\begin{equation}
\Gamma_\mu={1\over 2}[u^\dagger ,u]-{i\over
2}u^\dagger(v_\mu+a_\mu)u-{i\over 2}u(v_\mu-a_\mu)u^\dagger
\end{equation}
Even at this level this simple form has a number of phenomenological
successes:
\begin{itemize}

\item [i)] The Goldberger-Treiman relation---$Mg_A=F_\pi g_{\pi NN}$---is
valid to the accuracy of a few percent.\cite{12}

\item [ii)] In muon capture the prediction
\begin{equation}
{g_P\over g_A}={2Mm_\mu\over m_\pi^2+0.9m_\mu^2}=7.0
\end{equation}
is verified (although the recent TRIUMF radiative muon capture data
seems to be at odds with this prediction.)\cite{13}

\item [iii)] The Kroll-Ruderman theorem
\begin{equation}
E_{0+}^{\pi^\pm N}=\pm{\sqrt{2}eg_A\over 8\pi F_\pi}
\end{equation}
has recently been verified in threshold charged pion photoproduction.\cite{14}

\end{itemize}
One clearly wants to go to loop level in order to enforce
unitarity and to provide a stringent test of these ideas.  Loop
diagrams introduce divergences, but these can, of course, be absorbed
by empirical counterterm contributions, just as in the mesonic sector.
However, there is one complication which arises for baryons.  If one
simply uses Eq. \ref{eq:pp} and calculates the associated loop
diagrams using relativistic perturbation theory, then one finds that a
given loop diagram contributes to many different orders in the
mass/momentum expansion (generically denotes by $p$)---one does not
have consistent power counting.  In order to remedy this problem, 
a Foldy-Wouthuysen transformation is performed,\cite{15} so that one ends up with an
expansion both in $p/\Lambda_\chi$, where $\Lambda_\chi$ is the chiral
scale, as well as in $p/M$, where M is the baryon mass.  Of course,
the classic procedure of Foldy and Wouthuysen is performed in the
Hamiltonian picture, which it is more convenient for our purposes to
utilize a Lagrangian framework, as we now show.

The procedure of Foldy and Wouthuysen is one where coupling between
the large and small components of the baryon wavefunction is
eliminated.\cite{15}  Thus writing, for example,

\begin{equation}
\Psi(x,t)\equiv e^{-iMt}\left(
\begin{array}{l}
N(x,t)\\
H(x,t)
\end{array}\right)
\end{equation}
We can write the Lagrangian of the system in terms of upper, lower
components $N,H$ as
\begin{equation}
{\cal L}=\bar{N}AN+\bar{H}BN+\bar{N}\gamma_0B^\dagger\gamma_0H+\bar{H}CH
\end{equation}
This form can be diagonalized via the definition
\begin{equation}
H'=H+C^{-1}BN
\end{equation}
which yields
\begin{equation}
{\cal L}=\bar{N}(A-\gamma_0B^\dagger\gamma_0C^{-1}B)N+\bar{H}'CH'
\end{equation}
The ``heavy'' components $H'$ can now be integrated out, yielding an
effective Lagrangian in terms of the ``light'' components $N$
\begin{eqnarray}
W&=&\int[dN][d\bar{N}][dH'][d\bar{H}']\exp i\int d^4x{\cal
L}(N,H')\nonumber\\
&=&{\rm const.}\times\int[dN][d\bar{N}]\exp i\int d^4x{\cal L}_{eff}(N)
\end{eqnarray}
Using this form an extensive series of calculations involving low
energy nucleon electromagnetic interactions have been carried out by
the group of Bernard, Kaiser and Meissner (BKM).\cite{16}  In particular, in the case
of nucleon Compton scattering at one loop---${\cal O}(p^3)$---order
they find results
\begin{equation}
\alpha_E^p=\alpha_E^n=10\beta_M^p=10\beta_M^n=
{5e^2g_A^2\over 384\pi^2F_\pi^2m_\pi}=12.2
\end{equation}
which are in remarkable agreement with experiment!  As a reality
check, it should be
noted, however, that in the case of the spin polarizability the one
loop prediction
\begin{equation}
\gamma={e^2g_A^2\over 96\pi^3F_\pi^2m_\pi^2}=4.5\times 10^{-4}{\rm fm}^4
\end{equation}
is opposite in sign to the value Eq. \ref{eq:qq} given by the sum rule.
Also, BKM together with Schmidt 
have extended their calculation to ${\cal O}(p^4)$, yielding\cite{17}
\begin{eqnarray}
\alpha_E^p&=&10.5\pm 2.0\qquad \alpha_E^n=13.4\pm 1.5\nonumber\\
\beta_M^p&=&3.5\pm 3.6\qquad \beta_M^n=7.8\pm 3.6
\end{eqnarray}
The error bars here are associated with the feature that BKM
estimated the contribution of appropriate resonances, such as the
$\Delta(1232)$ by integrating them out and including only their
contribution to counterterms.  Also there exist various uncertainties
associated with higher order loop contributions.

However, while treating resonant contributions in this way is
satisfactory in general, it is not so clear that this is appropriate
for the $\Delta(1232)$ due to its strong coupling and its proximity to
the nucleon.  Thus I, together with Thomas Hemmert and Joachim Kambor, 
have developed a scheme whereby a chiral expansion can be performed
consistently with the $\Delta$ included as a specific degree of
freedom.\cite{18}  The difficulty here is that the Rarita-Schwinger
representation of the spin 3/2 field has too many degrees of
freedom---in addition to the desired spin 3/2 piece there exist {\it two}
independent spin 1/2 sectors.  Also, for each sector there is a large and
a small component.  We handled this problem by using
projection operators to identify each piece.  Thus we were able to
represent the Rarita-Schwinger field as a six component ``spinor''
\begin{equation}
\Psi^\mu=e^{-iMt}\left(
\begin{array}{l}
\Delta^\mu_{3\over 2}\\
H^\mu_{3\over 2}\\
\ell^{1\mu}_{1\over 2}\\
h^{1\mu}_{1\over 2}\\
\ell^{2\mu}_{1\over 2}\\
h^{2\mu}_{1\over 2}
\end{array}
\right)
\end{equation}
Then one writes the spin 3/2 Lagrangian in terms of these six fields
and integrates out all but the desired $\Delta^\mu_{3\over 2}$
component.  There are a few technically challenging features such as
taking the inverse of a $5\times 5$ matrix, but basically this is just
a repeat of what was done in the case of the Dirac---spin 1/2---field.
What results is a generalized chiral expansion in terms of
``small'' quantities $p, m_\pi$ and $\Delta\equiv m_\Delta-m_N$, which
we denote generically by $\epsilon$.  There exist in general two kinds
of additional contributions to the usual heavy baryon results.  One is
from the diagrams wherein the $\Delta(1232)$ appears as a simple pole,
while the other is where the $\Delta$ contributes as part of a loop
term.  Using couplings determined empirically we find in this way at
${\cal O}(\epsilon^3)$\cite{19}
\begin{eqnarray}
\alpha_E&=&12.2({\rm N-pole})+0(\Delta-{\rm pole})+4.2(\Delta-{\rm
loop})\nonumber\\
\beta_M&=&1.2({\rm N-loop})+7.2(\Delta-{\rm pole})+0.7(\Delta-{\rm
loop})\nonumber\\
\gamma&=&4.6({\rm N-loop})-2.4(\Delta-{\rm pole})-0.2(\Delta-{\rm loop})
\end{eqnarray}
Obviously in the case of the electric or magnetic polarizabilities the
$\Delta(1232)$ contributions are large and destroy agreement with
experiment, while in the case of the spin polarizability the
corrections are significant and in the right direction but are not
large enough to bring about agreement with the sum rule value.  In any
case it is clear that ${\cal O}(\epsilon^4)$ calculations are
abolutely necessary, and these are underway.

Before leaving this section it should be noted that a recent analysis by the
LEGS group of the world set of proton Compton scattering data has
produced a number for the {\it backward} spin-polarizability $\gamma_\pi$\cite{86}
\begin{equation}
\gamma_\pi=(-28.0\pm 2.8\pm 2.5)\times 10^{-4}{\rm fm}^4
\end{equation}
This quantity is simply the $180^\circ$-scattering analog of the usual
spin polarizability.  A direct measurement, of course, requires a
polarized beam {\it and} target.  However, the angular distribution of
the unpolarized cross section is also sensitive to $\gamma_\pi$ and
that is how it was extracted.  On the theoretical side the backward
spin-polarizability is dominated by the anomaly contribution from the
pion pole term, which alone yields a predicted effect
\begin{equation}
\gamma_\pi^{anomaly}=-44\times 10^{-4}{\rm fm}^4
\end{equation}
which is an order of magnitude larger than the size of its forward
scattering analog---$\gamma$---to which the pion pole diagram does not
contribute.  However, in addition to the pole term there are
additional contributions from the usual $N,\Delta$ loop and pole
diagrams, which tend to make the predicted value for $\gamma_\pi$
somewhat smaller in magnitude but in basic agreement with the measured
number\cite{87}
\begin{equation}
\gamma_\pi^{\rm theo}=[-44({\rm anomaly)}+4.6(N-{\rm
loop})+2.4(\Delta-{\rm pole})-0.2(\Delta-{\rm loop})]\times
10^{-4} {\rm fm}^4
\end{equation}
However, in this case not only additional theoretical work extending
these results to ${\cal O}(\epsilon^4)$ will be required, but also
direct experimental measurement using polarized beam and target in
order to have real confidence in the measured number.   Such
experiments can be expected at MAMI, LEGS, as well as at the free
electron laser backscattering facility now under development at Duke.

\section{Virtual Compton Scattering}

A new frontier in this area is represented by the subject of virtual
Compton scattering (VCS).  There are two different ways in which this can be
manifested.  One is to have a real incident photon but for the final
photon to fragment into a Dalitz pair.\cite{88}  This corresponds to positive
$q^2$ and will not be discussed here.  Rather we concentrate on the
case that the initial photon is produced in an electron scattering
process, with $q^2<0$, but scatters from a target to a real final
state photon.  This sort of process leads to probes of nucleon
structure via so called generalized polarizabilities, as we show below,
and has generated approved experimental programs at MAMI, CEBAF and
BATES.  A significant theoretical interest has also developed in VCS,
with many papers already having appeared.\cite{89}

On the experimental side VCS offers a significant advantage over its
usual Compton counterpart in that event rates possible with virtual
photons are much enhanced.  However, there is at the same time an associated cost
in that the desired process is hidden in general behind a huge
background due to Bethe-Heitler scattering, wherein the source of the
final photon is simply bremsstrahlung from either the initial or final
state electron.  On the experimental side this means that typical
measurements, which take place in ``parallel kinematics'' ({\it i.e.}
zero angle $\phi$ between
the lepton scattering plane and the hadronic scattering plane)
must extract
the interesting signal from a huge but calculable background flux.
The finite size of the detectors at MAMI, which extend out to 
$\phi=\pm 22^\circ$, ameliorates some of this effect, but it remains a
significant problem---the calculated generalized polarizability
``signal'' is only a very small component in a large Bethe-Heitler
``background.''\cite{90}  On the other hand, using the new and
``portable'' OOPS detectors, the BATES experiment will be able to
employ perpendicular kinematics---$\phi=90^\circ$---which
puts the signal and background on a nearly even footing.\cite{91}  In any case,
as theorists we can easily isolate the Bethe-Heitler from the VCS
signal---
\begin{equation}
{\rm Amp}^{\rm tot}_{\rm VCS}={\rm Amp}^{\rm
Bethe-Heitler}+ie^2(\vec{\epsilon}_T\cdot\vec{M}_T+{q^2\over \omega^2}
\epsilon_zM_z)
\end{equation}
where we have separated the VCS component into longitudinal and
transverse components and have used the shorthand
\begin{equation}
\epsilon_\mu=\bar{u}_{e'}\gamma_\mu u_e/q^2
\end{equation}
It is then straightforward to identify twelve---four longitudinal 
and eight transverse---independent structure functions, 
\begin{eqnarray}
\vec{\epsilon}_T\cdot\vec{M}_T&=&
\vec{\epsilon}'\cdot\vec{\epsilon}_TA_1+
\vec{\epsilon}'\cdot\hat{q}\vec{\epsilon}_T\cdot\hat{q}'A_2\nonumber\\
&+&i\vec{\sigma}\cdot\vec{\epsilon}'\times\vec{\epsilon}_TA_3
+i\vec{\sigma}\cdot\hat{q}'\times\hat{q}\vec{\epsilon}'\cdot\vec{\epsilon}_T
A_4\nonumber\\
&+&i\vec{\sigma}\cdot\vec{\epsilon}'\times\hat{q}\vec{\epsilon}_T\cdot\hat{q}'A_5
-i\vec{\sigma}\cdot\vec{\epsilon}'\times\hat{q}'\vec{\epsilon}_T\cdot\hat{q}'A_6\nonumber\\
&-&i\vec{\sigma}\cdot\vec{\epsilon}_T\times\hat{q}'\vec{\epsilon}'\cdot\hat{q}A_7
-i\vec{\sigma}\cdot\vec{\epsilon}_T\times\hat{q}\vec{\epsilon}'\cdot\hat{q}A_8\nonumber\\
M_z&=&\vec{\epsilon}'\cdot\hat{q}A_9+i\vec{\sigma}\cdot\hat{q}'\times\hat{q}
\vec{\epsilon}'\cdot\hat{q}A_{10}\nonumber\\
&+&i\vec{\sigma}\cdot\vec{\epsilon}'\times\hat{q}A_{11}
+i\vec{\sigma}\cdot\vec{\epsilon}'\times\hat{q}'A_{12}
\end{eqnarray}
For each structure function one expands
\begin{equation}
A_i=A_i^{\rm Born}+\mbox{generalized polarizabilities}
\end{equation}
where here $A_i^{\rm Born}$ signifies the nucleon pole diagrams with
on-shell form factors, while the generalized polarizabilities have
been defined by Guichon et al. in the low-energy approximation of 
including terms only up to linear
order in the real photon energy $\omega'$.\cite{20}  

\begin{table}
\begin{center}
\begin{tabular}{c|c|c|c}
Multipoles & S & Inter. St. & ($\rho'L',\rho L)S$\\ \hline
$L1\times L1$ & 0,1 & ${1\over 2}^-,{3\over 2}^-$ & $P^{(01,01)S}$\\
$L1\times E1$ & 0,1 & ${1\over 2}^-,{3\over 2}^-$ & $\hat{P}^{(01,1)S}$\\
$M1\times M1$ & 0,1 & ${1\over 2}^+,{3\over 2}^+$ & $P^{(11,11)S}$\\
$L2\times M2$ & 1   & ${3\over 2}^-$ & $P^{(01,12)1}$ \\
$M1\times L2$ & 1 & ${3\over 2}^+$ & $P^{(11,02)1}$\\
$M1\times L0$ & 1 & ${1\over 2}^+$ & $P^{(11,00)1}$\\
$M1\times E2$ & 1 & ${1\over 2}^+$ & $\hat{P}^{(11,2)1}$
\end{tabular} \newline
\caption{Generalized polarizabilities as defined by Guichon et al.\cite{20}}
\end{center}
\end{table}

As summarized in Table 1, there exist ten such terms---three of 
which are spin-independent
and seven requiring polarization for their measurement.  However, it
was subsequently demonstrated by Drechsel et al. that a consistent
treatment of crossing symmetry and charge conjugation invariance
yields four additional constraints---one for S=0 and three for S=1.\cite{21}
In the spin-independent case one can then eliminate $\hat{P}^{(01,1)0}$
and write everything in terms of just the two {\it generalized}
polarizabilities
\begin{eqnarray}
\alpha_E(\bar{q})&=&-{e^2\over 4\pi}\sqrt{3\over
2}P^{(01,01)0}(\bar{q})\nonumber\\
\beta_M(\bar{q})&=&-{e^2\over 4\pi}\sqrt{3\over 8}P^{(11,11)0}(\bar{q})\label{eq:gg}
\end{eqnarray}
which reduce to the usual quantities in the real photon limit
$\bar{q}\rightarrow 0$.  The meaning of these quantities is also
clear.  When one applies an electric or magnetic field to a charged
system the induced electric or magnetic dipole moments are in general
functions of position, whose Fourier transform in $\bar{q}$ are just
the generalized polarizabilities given above.  
In the spin-dependent case it is not so clear which generalized
spin-polarizbilities to eliminate, but in any case charge conjugation
invariance implies that there exist only four independent
$\bar{q}$-dependent quantities.

In the first---unpolarized---experiments what will be measured are
three independent combinations
\begin{eqnarray}
P_{LL}(\bar{q})&=&-2\sqrt{6}MG_E(Q_0^2)P^{(01,01)0}(\bar{q})\nonumber\\
P_{TT}(\bar{q})&=&{3\over
2}G_M(Q_0^2)\left[2\omega_0P^{(01,01)1}(\bar{q})
+\sqrt{2}\bar{q}^2(P^{(10,12)1}(\bar{q})+\sqrt{3}\hat{P}^{(01,1)1}(\bar{q}))\right]
\nonumber\\
P_{LT}(\bar{q})&=&\sqrt{3\over 2}{M\bar{q}\over \sqrt{Q_0^2}}
G_E(Q_0^2)P^{(11,11)0}(\bar{q})+{\sqrt{3}\sqrt{Q_0^2}\over
2\bar{q}}G_M(Q_0^2)\nonumber\\
&\times&\left[P^{(11,00)1}(\bar{q})+{\bar{q}^2\over \sqrt{2}}
P^{(11,02)1}(\bar{q})\right]
\end{eqnarray}
The leading terms here are the LL and LT pieces so that one's initial
sensitivity will be to the electric and magnetic generalized
polarizabilities given in Eq. \ref{eq:gg}.  A particularly interesting
test here will be the measurement of the magnetic polarizability 
$\beta_M(\bar{q})$, for which loop effects in heavy baryon chiral
perturbation theory predict a temporary rise (!) at low $\bar{q}^2$ in
contradistinction to quark models which predict a steady decrease due
to form factor effects.\cite{53}  The news here then is good
and bad.  On the one hand the technique of virtual Compton scattering
offers a new and potentially high resolution probe of nucleon
structure.  On the
other hand the new information is available of significant background
due to Bethe-Heitler and nucleon Born diagrams and very high precision
experiments will be required in order to harvest this potentially rich
crop of data.

\section{Conclusions}

During the past decade Compton scattering has become an important tool
for the probing of hadron structure, and this will no doubt continue
into the new millenium.  Indeed there remain significant challenges
for both experimentalists and theorists in this regard.  In the former
case, the challenges will be to improve upon existing measurements of
both proton and (especially) neutron polarizabilities, as well as to
utilize polarized beam and target technology to provide new
spin-polarizability measurements.  Experiments yielding polarized
photoabsorption cross sections should also become available in order to
test the various sum rule predictions.  Virtual Compton scattering
programs at the electron machines will provide a rich lode of new
generalized polarizability information.  On the theoretical side the
challenge will be to understand this rich trove of information.  One
frontier is to provide chiral calculations at ${\cal O}(\epsilon^4)$.
A second is to relate this information to dispersive
approaches which provide the high mass contributions to sum rules for
these quantities.  Finally, I personally would like to develop also a
{\it physical}
understanding for the meaning of each of these generalized
polarizabilities so that we can communicate with colleagues what the
excitement of these measurements really means.

\end{document}